\documentclass[10pt,a4paper]{article}
\usepackage[utf8]{inputenc}
\usepackage[english]{babel}
\usepackage{amsmath}
\usepackage{amsfonts}
\usepackage{amssymb}
\usepackage{makeidx}
\usepackage{graphicx}
\usepackage{lmodern}
\usepackage{kpfonts}
\usepackage{natbib}
\usepackage{ctable}
\usepackage[left=2cm,right=2cm,top=2cm,bottom=2cm]{geometry}

\title{Estimating abilities with an Elo-informed growth model}
\author{Karl Sigfrid$^{1}$, Ellinor Fackle-Fornius$^{1}$, Frank Miller$^{2}$\\[2mm]
$^1$Department of Statistics, Stockholm University, Sweden\\
$\mbox{}^2$Department of Computer and Information Science, Link\"oping University, Sweden}
\date{\today}

\begin{document}

\maketitle

\begin{abstract}
An intelligent tutoring system (ITS) aims to provide instructions and exercises tailored to the ability of a student. To do this, the ITS needs to estimate the ability based on student input. Rather than including frequent full-scale tests to update our ability estimate, we want to base estimates on the outcomes of practice exercises that are part of the learning process. A challenge with this approach is that the ability changes as the student learns, which makes traditional item response theory (IRT) models inappropriate. Most IRT models estimate an ability based on a test result, and assume that the ability is constant throughout a test.

We review some existing methods for measuring abilities that change throughout the measurement period, and propose a new method which we call the Elo-informed growth model. This method assumes that the abilities for a group of respondents who are all in the same stage of the learning process follow a distribution that can be estimated. The method does not assume a particular shape of the growth curve. It performs better than the standard Elo algorithm when the measured outcomes are far apart in time, or when the ability change is rapid.
\end{abstract}

\section{Introduction}

Smart learning with the help of information and communication technologies is a field that has received increasing interest. A central topic in this area is learning analytics, i.e. the use of student generated data to support the learning process. With the development of technologies such as virtual reality, the trend may go beyond today's web-based applications \citep{chen_past_2021}.

A meta-analysis of 50 evaluations of Intelligent Tutoring Systems (ITS) found that students who received intelligent tutoring outperformed student who did not receive this help. In 39 of the 50 studies, the improvement was considered substantial \citep{kulik_effectiveness_2016}. The promise of improved learning outcomes through ITS makes the field likely to keep expanding.

ITS is a broad term, which can describe any computer program that uses some type of intelligence to support learning. A key function is the ability of the tutoring system to adapt to the needs of individual students. Thus, an ITS tries to identify student characteristics, and given these characteristics it sets a course of action to achieve the learning goals \citep{mousavinasab_intelligent_2021}.

The student characteristic of interest in this paper is the student's ability, i.e. the skill of a student on the topic that the ITS is teaching. Knowledge about the skill level can be instrumental in determining the appropriate difficulty of instructions and practice exercises. When the diagnostics and the remedies are provided by an ITS, we call it adaptive learning, which has as its goal to teach each student at the right level. Teaching at the right level is a concept that can be used also in non-digital settings, e.g. by grouping students according to ability and then teaching each group separately. Projects that implement teaching at the right level have in some studies shown positive effects \citep{duflo_peer_2011}.

A central component in adaptive learning is a method to estimate the ability of a student. There are several well-established methods to estimate an ability through testing, such as Item Response Theory (IRT). However, an assumption for most IRT models is that the ability is constant throughout the test. The respondent is expected to have the same ability when the test ends as when the test started.

In an adaptive setting, where instructions and exercises are provided as part of the learning process, we expect a growing ability. In some circumstances, as when a respondent has not engaged in learning for a period of time, we may also see a decline in ability. The aim of this paper is to describe some available methods for tracking a changing ability, and to propose a new method that achieves this with good accuracy under challenging circumstances, such as rapid ability growth. Compared to some of the available methods, our proposed method does not make any assumption about the shape of the ability growth. The method is flexible in the sense that it allows for the students to receive personalized learning items at different time intervals throughout the process. A change in ability between successive items may either be due to separate learning activities or that an ability growth results a direct effect of answering the item. 

We will here use the term \textit{iteration} to refer to the position in a sequence of responses. For instance, the first iteration is the time point of a respondent's first response. Two respondents are both at the $t$:th iteration when they answer their $t$:th item, regardless of whether these responses are far apart in chronological time or not. 

A challenge in measuring an ability that changes between responses is that we only have one observation that reflect their ability at each iteration. If the most recent item was administered at iteration $t$, then the previous items were administered at iterations $t-1$, $t-2$, etc. If the respondent's ability was different at these earlier iterations, then the outcomes of the earlier responses are not products of the current ability.

Our approach to solve the problem is to view respondents who are at the same iteration as a group. Whereas we cannot use a single outcome to reliably estimate the ability of one respondent at iteration $t$, we can use the outcomes from a group of respondents at iteration $t$ to estimate the ability distribution of that group. Making the assumption that the ability distribution at each iteration is normal, we fit a normal distribution that represents the abilities of the group of respondents.

Further, each respondent is assigned a ranking. This ranking, together with the normal distribution that we fitted, determines the ability estimate of the respondent at that iteration. To rank respondents we use the Elo algorithm \citep{elo_rating_1978}. Note that whereas we use the Elo algorithm to determine the ranking of the respondents, we do not use Elo estimate as our final ability estimate. By instead combining Elo ranking with a normal assumption at the group level, we get an estimate that is more robust to rapid ability changes.

As with unidimensional IRT we assume that we want to measure a single ability, and that we have test items that measure this ability. We also assume that the items in our item bank have known difficulty. \cite{wauters_item_2012} describes it as common practice in IRT to obtain a pool of items that have been pre-calibrated. The pre-calibration can be achieved by administering the items to a large group of respondents in a non-adaptive setting. The calibration of the items is outside the scope of this paper.

The paper is organized as follows. We start by describing the test design we consider in Section 2. Section 3 contains a review of alternative methods to track changing abilities and describe their merits and limitations. Further, Section 3 includes the details of our proposed method. We demonstrate the method on a real data set which we present in Section 4. In Section 5 we compare the results of our proposed method to some selected alternative methods in terms of accuracy and fitting speed. The paper ends with a discussion and summary of the conclusions.

\section{Test design}

In our test design we have a respondent who participates in a course on a specific topic. The respondent's ability with regards to the topic can be measured on a unidimensional scale. The item difficulties are known. Throughout the course, the respondent answers a sequence of test items, which may be in the form of practice exercises. We refer to each response in the sequence as an iteration. Iteration 1 is thus the response to the first item in a respondent's sequence. The chronological time between two iterations does not need to be constant, but can vary widely within the sequence. It can also vary widely between respondents. Even in the case where two items are answered immediately after each other, we view them as two different iterations in the sequence.

After each iteration, the respondent's ability estimate is updated. The goal is to continuously track the ability as it changes. The ability at the current iteration will typically differ from the ability at previous iterations. The course may or may not be adaptive. If it is adaptive, instructions and test items will be at a level matching the respondent's estimated ability.

\subsection{Notation}

The majority of the methods in the next section aim to estimate $\theta_{j,t}$, which is the ability of a respondent $j$ at an iteration $t$. Iteration $t$ refers to position $t$ in the sequence of items. The difficulty of an item that respondent $j$ encounters at iteration $t$ is denoted $d_{j,t}$. The item that one respondent encounters at iteration $t$ may or may not be the same as the item that another respondent encounters at iteration $t$. $Y_{j,t}$ is a Bernoulli distributed random variable with success-probability $p_{j,t}$ that takes the value 1 if respondent $j$ answers the item at iteration $t$ correctly. Otherwise, it takes the value 0.

\section{Methods}

\subsection{Methods overview}

There are several types of models frequently used to estimate abilities based on item responses. Item response theory models, such as the Rasch model or variants that incorporate additional parameters, are standard for estimating an ability from a test outcome. An assumption in most IRT models is that the respondent's ability is the same throughout the test. For our purposes, we are interested in models that can account for changes in ability from one test item to another. We will here describe existing models that aim to do this. Some of these methods we will include in our models comparison in the results sections. Other methods in this section will not be included in our models comparison, but are nevertheless described here to broaden the overview of methods that have been used for the purpose of tracking ability growth.

\subsubsection{Learning Factors Analysis}

Learning Factors Analysis (LFA) extends the Rasch model \citep{cen_learning_2006}. Instead of assuming a constant ability, LFA models ability as a linear function of the number of previously answered items. A respondent has an individual intercept that represents the starting ability, and the ability improves when the respondent answers test items. Further, an item may involve multiple knowledge components, i.e. it may require multiple skills that each has its own difficulty.

LFA models the logarithmized odds of a correct answer from respondent $j$ at time point $t$ as
\begin{equation}
\label{eqn:LFA}
\ln\left( \cfrac{p_{j,t}}{1-p_{j,t}} \right) = \alpha_j  + \sum_{c \in \Omega}(\gamma_{c} m_{j, t, c} - \beta_{j,t,c}),
\end{equation}
where $\alpha_j$ is an intercept for respondent $j$. A higher value of $\alpha_j$ implies a higher ability at the time of the first test item. The index $c$ represents a knowledge component, and $\Omega$ is the set of knowledge components required to solve the item. Further, $\beta_{j,t,c}$ is the difficulty of the item theat respondent $j$ encounters at time $t$ with respect to component $c$, $m_{j,t,c}$ is the number of items that respondent $j$ has already attempted at time $t$ which involve knowledge component $c$, and $\gamma_c$ is the ability growth that we expect from each such attempt. The model assumes linear ability growth and does not distinguish between successful attempts and unsuccessful attempts.

\subsubsection{Performance Factors Analysis}

Performance Factors Analysis (PFA) has been developed out of LFA to make it useful in adaptive settings where we want to track the ability of a student \citep{pavlik_performance_2009}. A main difference between PFA and FLA is that PFA treats correct and incorrect answers differently. Also, the individual intercept $\alpha_j$ has been removed from the model.

PFA models the logarithmized odds of a correct response as
\begin{equation}
\label{eqn:PFA}
\ln\left( \cfrac{p_{j,t}}{1-p_{j,t}} \right) = \sum_{c \in \Omega}(\gamma_{c} s_{j,t,c} + \rho_c f_{j,t,c} - \beta_{j,t,c}),
\end{equation}
where $s_{j,t,c}$ is the number of correct answers that respondent $j$ has given to items that involve knowledge component $c$ up to time $t$. $\gamma_c$ determines how much the ability increases for each correct answer. The parameter $f_{j,t,c}$ is the number incorrect responses, and $\rho_c$ the effect of each incorrect response on the ability estimate. The parameter $\beta_{j,t,c}$ represents the difficulty of the item at time $t$ with respect to knowledge component $c$. \cite{pavlik_performance_2009} suggest fitting the model parameters with a Maximum Likelihood procedure.

Whereas PFA extracts more information from the data than LFA, as is distinguishes successfull attempts from failed attempt, it does not account for each item's individual difficulty. Two items that involve the same knowledge components are considered equally difficult.

\subsubsection{Knowledge tracing}

In Knowledge tracing, ability is defined as the probability that a respondent masters a topic \citep{atkinson_ingredients_1972, corbett_knowledge_1994}. What we here refer to as Knowledge Tracing (KT) is also called Bayesian Knowledge Tracing or Standard Bayesian Knowledge Tracing. Many varieties of KT have been developed with additional parameters, to accommodate factors like individual learning speed or the possibility of a respondent forgetting a previously mastered topic\citep{abdelrahman_knowledge_2023, shen_survey_2024}. After each response we estimate the probability that a respondent $j$ masters the topic conditioned on the previous estimate and on whether the last response was correct or not. Depending on the last outcome, we use either the formula
\begin{equation}
\label{eqn:BKT1}
P(L_{j,t-1}|\text{correct}_t) = \cfrac{P(L_{j,t-1}) \cdot (1 - P(S))}{P(L_{j,t-1}) \cdot (1 - P(S)) + (1 - P(L_{j,t-1})) \cdot P(G)}
\end{equation}
or
\begin{equation}
\label{eqn:BKT2}
P(L_{j,t-1}|\text{incorrect}_t) = \cfrac{P(L_{j,t-1}) \cdot P(S)}{P(L_{j,t-1}) \cdot P(S) + (1 - P(L_{j,t-1})) \cdot (1 - P(G))}.
\end{equation}
To estimate the ability, we need 4 parameters.\begin{itemize}
\item $P(L_{0})$ is the initial probability that a respondent masters the topic.
\item $P(S)$ is the probability of a slip, i.e. that a respondent who masters the topic still gets the answer wrong.
\item $P(G)$ is the probability that a respondent who does not master the topic guesses the correct answer.
\item $P(T)$ is the probability that a respondent learns the topic, i.e. that a student who did not master the topic when answering item $t-1$ masters it when answering item $t$.
\end{itemize}

The probabilities of a slip, of a correct guess or of learning the topic are here assumed to be the same for all subjects.

Formula \ref{eqn:BKT1} says you can answer an item correctly in two ways. One way is to master the topic and not slip, and the other way is to not master the topic but make a correct guess. Likewise there are two ways to fail, as expressed in formula \ref{eqn:BKT2}. One way is to master the topic but slip, and the other is to not master the topic and not make a correct guess.

Formula \ref{eqn:BKT1} or \ref{eqn:BKT2} gives us an estimate of the probability that the respondent mastered the topic when answering item $t-1$, i.e. at the iteration before item $t$. To estimate the probability that the respondent mastered the topic when answering item $t$, we add the step
\begin{equation}
\label{eqn:BKT3}
P(L_{j,t}|\text{outcome}_t) = P(L_{j,t-1}|\text{outcome}_t) + (1 - P(L_{j,t-1}|\text{outcome}_t)) \cdot P(T).
\end{equation}
Formula \ref{eqn:BKT3} say that two scenarios imply mastery. One is that respondent $j$ already mastered the topic when answering item $t-1$, and the other is that respondent $j$ learned it after answering item $t-1$. Like PFA, KT does not account for items with different difficulty in a univariate setting. Instead, the model views the probability of a correct answer as equally large for all items that measure the same skill.

\subsubsection{Elo}

The Elo rating system was developed to rate chess players \citep{elo_rating_1978}. It has since become a widespread method used for many different purposes, such as rating computer game players and measuring student ability \citep{antal_use_2013, pelanek_applications_2016}.

Whereas the probability for a win in a chess game is typically written in the form
\begin{equation}
\label{eqn:elo1}
p=\cfrac{10^{(R_1 - R_2)/400}}{1 + 10^{(R_1 - R_2)/400}},
\end{equation}
with $R_1$ and $R_2$ being the player's and opponent's rating, respectively,
we can transform the parameters, which allows us to use the form
\begin{equation*}
p_{j,t}=\cfrac{e^{\theta_{j,t} - d_{j,t}}}{1 + e^{\theta_{j,t} - d_{j,t}}},
\end{equation*}
or, equivalently,
\begin{equation}
\label{eqn:elo2}
\ln\left( \cfrac{p_{j,t}}{1-p_{j,t}} \right)=\theta_{j,t} - d_{j,t},
\end{equation}
which is the formula for calculating the probability of a correct answer in the Rasch model. Note that ELO chess ratings are sometimes not updated after each game but after tournaments. Here for our educational setting, we allow for updates after each item. We could have assumed that when multiple items are answered at one occasion, the responses are manifestations of one and the same ability. However, updating the ability after each response is justified if we expect that responding to an item is in itself a learning activity.

Just as KT, the Elo rating can be viewed as a Markov chain. The probability of a win, i.e. a correct answer to an item in an educational setting, is a function of the current ability and the difficulty of the item. After an item has been answered, the Elo rating updates with the formula
\begin{equation}
\label{eqn:elo3}
\hat{\theta}_{j,t} = \hat{\theta}_{j,t-1} + K(y_{j,t} - p_{j,t}),
\end{equation}
where $y_{j,t}=1$ if the outcome is a correct answer, and otherwise $y_{j,t}=0$. The rating changes more if there is a larger difference between the outcome and the expected outcome. The parameter $K$ regulates the step size. In chess practise, $K$ can depend on the rating of the player or the age.

The Elo method is computationally inexpensive, and it also has the advantage of not assuming a certain shape of the growth curve. Several extensions of the Elo system have been suggested \citep{pelanek_applications_2016}. One such extension is Performance Factor Analysis Extended (PFAE), where $K$ is different depending on whether the response is correct or incorrect \citep{papousek_adaptive_2014}. Other extensions replace the constant $K$ with a $K$ that changes as a function of the number of previous answers. Typically, the value of $K$ shrinks as the number of previous answers increases. Thus the process starts with larger initial steps to find the neighborhood of the true ability, and then transitions to fine tuning \citep{niznan_student_2015}.

\subsubsection{Glicko}

The Glicko rating system was developed to improve the Elo rating system \citep{glickman_parameter_1999}. Whereas the Elo method produces point estimates of abilities, the Glicko method produces estimates in the form of probability distributions. Each player has an ability that follows a normal distribution with a mean and a standard deviation, both of which update after each encounter. The standard deviation is referred to as the Ratings Reliability (RD).

A player who has a lower RD is believed to have a more accuracte estimate. Therefore, a player with lower RD will see a smaller change in the estimated ability as the result of a game. When an opponent has a higher RD, this will result in a smaller change in the estimate. The reason for this is that we get less reliable information from a game against an opponent with a more uncertain ability. The RD of a player increases when time passes between games.

A later variety of the Glicko method called the Glicko-2 rating system introduced a volatility parameter. The volatility is low for players who perform at a consistent level and high for players whose performances are less consistent \citep{glickman_example_2012}. The Glicko rating system uses a Bayesian method of updating ability distributions.

\subsubsection{Generalized Linear Mixed Models}

A generalized linear mixed models (GLMM) is a generalized linear model that allows random effects in the linear term \citep{mcculloch_chapter_2003}. As a generalized linear model, it accomodates response data that is not normally distributed, such as a Bernoulli distributed variable that can represent correct and incorrect answers to an item \citep{mcculloch_chapter_2003}. With a Bernoulli distributed response variable $Y_{j,t}$, we get the linear function
\begin{equation}
\begin{aligned}
\label{eqn:glmm0}
\ln\left({\cfrac{p_{j,t}}{1-p_{j,t}}}\right) = \boldsymbol{x}_{j,t}^\intercal \boldsymbol{\beta} + \boldsymbol{z}_{j,t}^\intercal \boldsymbol{u}_j, \\
\boldsymbol{u}_j \sim {\rm Normal}(\boldsymbol{0}, \boldsymbol{\Sigma}),
\end{aligned}
\end{equation}
where $p_{j,t}$ is the probability that respondent $j$ answers the item at time point $t$ correctly, and $\ln\left(p_{j,t}/(1-p_{j,t})\right)$ are the log odds. 
The vectors $\boldsymbol{x}_{j,t}$ and $\boldsymbol{z}_{j,t}$ are known covariates, $\boldsymbol{\beta}$ an unknown parameter vector, $\boldsymbol{u}_j$ a random effects vector with mean $\boldsymbol{0}$ and a $p \times p$ variance-covariance matrix $\boldsymbol{\Sigma}$, where $p$ are the number of random effects associated with a respondent.

In general, $\boldsymbol{x}_{j,t}$ can be a vector of subject-specific covariates at time $t$. Here, we focus on the ability-change over time and let the vector depend on the time point or a stage in a learning process, $\boldsymbol{x}_t = \boldsymbol{x}_{j,t}$. Then, $\boldsymbol{x}_{t} \boldsymbol{\beta}$ represents the average log odds for respondents who are at time point $t$. If the time variable is a factor variable, then $\boldsymbol{x}_{t}$ is an $I$-dimensional vector of 0's and a 1 in dimension $t$ and $\boldsymbol{\beta}$ is a vector of parameters with $I$ elements, where $I$ is the number of time points at which we measure.

We can obtain estimate of the model parameters $\boldsymbol{\beta}$ and $\boldsymbol{\Sigma}$ by maximizing the likelihood

\begin{equation}
\label{eqn:glmm}
L(\boldsymbol{\beta}, \boldsymbol{\Sigma}|y_{j,t}) = \prod_j \int_{\boldsymbol{u}_j} \prod_t \cfrac{\exp(\boldsymbol{x}_{t}^\intercal \boldsymbol{\beta} + \boldsymbol{z}_{j,t}^\intercal \boldsymbol{u}_j)^{y_{j,t}}}{1 + \exp(\boldsymbol{x}_{t}^\intercal \boldsymbol{\beta} + \boldsymbol{z}_{j,t}^\intercal \boldsymbol{u}_j)} \cdot(2 \pi)^{-k/2} \det{(\boldsymbol{\Sigma})}^{-1/2} \exp(-\boldsymbol{u}_j^\intercal \boldsymbol{\Sigma}^{-1} \boldsymbol{u}_j/2) d\boldsymbol{u}_j.
\end{equation}

Note that $y_{j,t}$ is either 0 or 1. When it is 1, the logistic function gives the probability of a correct answer. When it is 0, the nominator becomes 1, and the logistic function gives the probability of an incorrect answer. 
The integral in \ref{eqn:glmm} is intractable which calls for methods for approximating the integral, before the likelihood can be maximized. Several strategies to handle the issue exist, differing with respect to accuracy, complexity and computational speed. A uniformly preferred method is not possible to appoint, since it depends on the study goals, design and model structure. For a thorough review of the different classes of methods, see \cite{tuerlinckx2006statistical}. These include direct maximization methods following either a non-stochastic (Gauss-Hermite quadrature or its improved adaptive version) or stochastic (e.g. Monte Carlo simulation) approximation of the integral, indirect maximization such as the Expectation-Maximization algorithm as well as approximations of the integrand in order to achieve tractability. Common integrand approximations are Laplace's method (involving a Taylor series expansion around the mode) and Quasi-likelihood approaches (building on different kinds of linearizations so that estimation methods developed for linear mixed models can be utilized). Laplace approximations have for example been demonstrated to be fast and efficient for simultaneous estimation of complex joint models of mixed data with multidimensional latent variables \citep{zhang2024fast}. We use the \texttt{glmer} function in the package \texttt{lme4} in R, which implements adaptive Gauss-Hermite. The default setting for the number of quadrature nodes is \texttt{nAGQ=1}, which is numerically equivalent to a Laplace approximation \citep{tuerlinckx2006statistical}.

When we estimate the model parameters $\boldsymbol{\beta}$ and $\sigma^2$, the  value of parameter $u_j$ is unknown, but we assume that it follows a normal distribution. For each respondent, the estimation of the model parameters requires us to integrate the likelihood over this normal distribution, which can result in a slow fitting process.

If we accept that the variance in ability between subjects is the same at every time point, then we only need a random intercept, $\boldsymbol{z}_{j,t}=1$. This allows us to replace $\boldsymbol{z}_{j,t}^\intercal \boldsymbol{u}_j$ by a scalar $u_j$ in the formula. The estimated variance of the intercept, i.e. the random variation between subjects, implies a random effect $u_j$ for each subject. With a random intercept as the only random effect, $\boldsymbol{\Sigma}$ becomes $\sigma^2$, which is a scalar representing the variance between respondents.

If we know that the item at time point $t$ has the difficulty $d_{t}$, which is measured on the same scale as the ability, this also needs to be included in the model. With these adjustments we get the model
\begin{equation}
\begin{aligned}
\label{eqn:glmm1}
\ln\left({\cfrac{p_{j,t}}{1-p_{j,t}}}\right) = \boldsymbol{x}_t^\intercal \boldsymbol{\beta} + u_j - d_{j,t}, \;\;\;\;\;\; p_{j,t} = P(y_{j,t} = 1),\\
Y_{j,t} \sim {\rm Bernoulli}(p_{j,t}),\\
u_j \sim {\rm Normal}(0, \sigma^2).
\end{aligned}
\end{equation}

The expression $\boldsymbol{x}_{t} \boldsymbol{\beta} + u_{j}$ is equivalent to the ability $\theta_{j}$ in a Rasch model. The variable $d_{j,t}$ is an offset, i.e. a variable associated with a parameter fixed to 1 \citep{roffset}. 
With a slight modification of the Maximum Likelihood expression from \citet{mcculloch_chapter_2003}, we can for the case with a univariate random effect obtain an estimate of the model parameters $\boldsymbol{\beta}$ and $\sigma^2$ by maximizing the expression
\begin{equation}
\label{eqn:glmm2}
L(\boldsymbol{\beta}, \sigma^2|y_{j,t}) = \prod_j \int_{u_j} \prod_t \cfrac{\exp(\boldsymbol{x}_{t}^\intercal \boldsymbol{\beta} + u_j - d_{j,t})^{y_{j,t}}}{1 + \exp(\boldsymbol{x}_{t}^\intercal \boldsymbol{\beta} + u_j - d_{j,t})} \cdot \cfrac{1}{\sqrt{2 \pi \sigma^2}} \exp(-u_j^2 / 2 \pi \sigma^2) du_j.
\end{equation}

Once we have estimated $\boldsymbol{\beta}$ and $\sigma^2$, we can estimate the random effect $u_j$ for each subject, i.e. how much a repondent's ability differs from the mean ability. As we have specified the model, this difference is assumed to be the same for all iterations.

We can estimate $u_j$ for each respondent individually by maximizing the likelihood
\begin{equation}
\label{eqn:glmm3}
L(u_j|y_{j,t}) = \prod_t \cfrac{\exp(\boldsymbol{x}_{t}^\intercal \boldsymbol{\beta} + u_j - d_{j,t})^{y_{j,t}}}{1 + \exp(\boldsymbol{x}_{t}^\intercal \boldsymbol{\beta} + u_j - d_{j,t})} \cdot \cfrac{1}{\sqrt{2 \pi \sigma^2}} \exp(-u_j^2 / 2 \pi \sigma^2).
\end{equation}

Unlike the estimation of the model parameters, this is a fast and simple optimization task. An alternative, more simplistic, method of estimating the ability of a respondent is to only use the fixed model parameters. This implies that we estimate a users ability to equal the mean ability for respondents at the same time point. While computationally inexpensive, this latter method discards valuable information about the individual respondents which results in less accuracy.

In the example shown here, each respondent is associated with a parameter $u_j$, which is the same for all time points. As a result, the variance in ability among respondents is constant over time. If there is reason to believe that the variance among respondents changes over time, we can replace $u_j$ with the original term $\boldsymbol{z}_{j,t}^\intercal \boldsymbol{u}_j$. For a simple linear change in the variance, $\boldsymbol{z}_{j,t}=(1, t)^\intercal$ and $\boldsymbol{u}_j$ includes a random intercept and a random slope associated with respondent $j$.

Depending on how a GLMM model is implemented it may or may not assume a specific shape of the ability growth curve. If we have a time variable that is numeric, then the model specified above will assume linear growth. However, if the time variable is a a factor variable, the mean ability at each time point is assigned its own parameter, which allows the growth curve to have any shape. GLMM can therefore be a viable option to track the ability development of a student over time also when we do not want to assume a shape of the growth curve.

\subsection{The Elo-informed growth model}

For every iteration $t$, i.e. for every time point when an item is answered, the respondent has an ability $\theta_{j,t}$. For instance, $\theta_{1,1}$ refers to the ability of respondent $j=1$ when this respondent answers their first item. We view all respondents who are at the same iteration, e.g. respondents who are answering their $t$:th item, as a group and assume that the abilities in this group follow a normal distribution. We do not assume that the item at some iteration $t$ is the same for all respondents, or that the learning activities between items are the same.

\begin{figure}
\begin{center}
\includegraphics[scale=0.7]{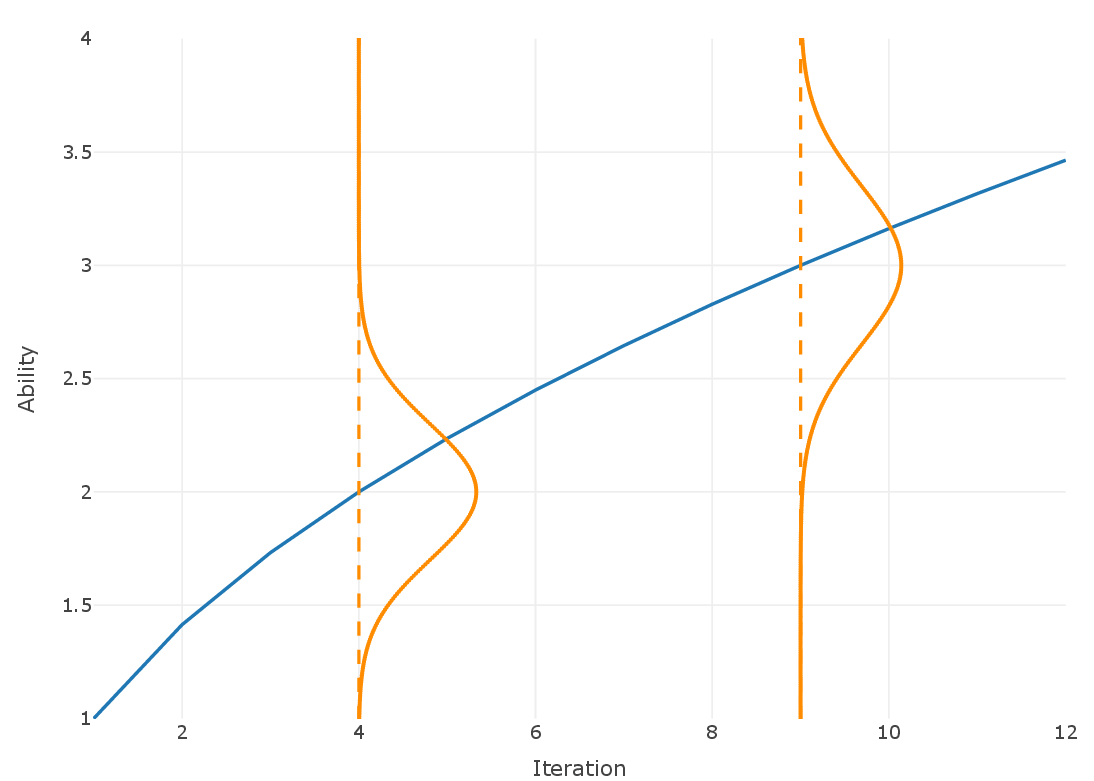}
\caption{We assume that a group of respondents at each iteration have abilities that follow a normal distribution.}
\label{fig:normal_distributions}
\end{center}
\end{figure}

The overall method is illustrated in figure \ref{fig:normal_distributions}, where we have plotted a mean growth curve and the ability distributions at iterations 4 and 9 respectively. We have a separate distribution for each iteration included in the model, and the growth curve goes through the distribution means.

The fitting of our model entails estimation of the the mean and the standard deviation of the normal distribution at each iteration. When we have a fitted model and want to estimate the ability of a new respondent at some iteration $t$, we view this respondent as member of the group associated with iteration $t$. To estimate where on the distribution this particular respondent is located, we use the Elo algorithm to rank the respondent against the the group that we used for the model fitting. We express the ranking as the proportion of respondents who have lower ability than the respondent of interest. Under our normal assumption, this proportion is associated with a z-value that tells us how many standard deviations above or below the group mean the respondent's ability is located. With the z-value and the fitted distribution, we have an estimate of the respondent's ability at iteration $t$.

Whereas the ranking of respondents in a group relies on the Elo algorithm, we do not assume that the Elo algorithm always gives a good ability estimate. If we did, the proposed method would be superfluous. However, this method does assume that the Elo estimates provide a sufficiently good ranking of the respondents. The goodness of the ranking can be measured, for instance, as the Spearman correlation, which we will return to in the results section.

\subsubsection{Fit the model}

Fitting the model involves the following steps:

\begin{enumerate}
\item Calculate a mean ability at iteration 1. This could be a Maximum Likelihood estimate under the assumption that all respondents have an equal starting ability.
\item Using the mean ability from the previous step as the starting ability for all respondents, calculate the Elo estimates for all respondents at all iterations.
\item For each respondent at each iteration, note the proportion $P_{j,t}$ of respondents who have a lower estimated Elo score.
\item Fit a normal distribution for each iteration. We do this by maximizing the likelihood function
\begin{equation}
\label{eqn:likelihood_function}
L(y_{j,t} | \mu_t, \sigma_t) = \prod_{\tau=1}^{n_{\tau}} \prod_{j=1}^{n_j}  w_{\tau} \cdot \cfrac{(e^{\hat{\theta}_{j,\tau}-d_{j,\tau}})^{y_{j,\tau}}}{1 + e^{\hat{\theta}_{j,\tau}-d_{j,\tau}}},\;\;\;\;\; \hat{\theta}_{j,\tau} = \hat{\mu}_{t} + \hat\sigma_{t} \cdot \Phi^{-1} \left( P_{j,\tau} \right).
\end{equation}

When we fit one iteration, we use data from all iterations. In the product, we use the index $\tau$ to represent all iterations, and the index $t$ to represent the iteration of the distribution that we are currently fitting. 
 
The parameter $d_{j,\tau}$ is the known difficulty of the item answered by respondent $j$ at iteration $\tau$. The response $y_{j,\tau}$ can have the value $y_{j,\tau}=0$ if the response is incorrect, or $y_{j,\tau}=1$ if the response is correct. If $y_{j,\tau}=0$, then the logistic function gives the probability of an incorrect answer. If $y_{j,\tau}=1$, then the logistic function gives the probability of a correct answer. In either case, the expression always gives the probability of the observed outcome.

$P_{j,\tau}$ is the proportion of respondents at iteration $\tau$ who have an Elo ranking lower than that of respondent $j$. The function $\Phi^{-1}(.)$ is the inverse of the cumulative standard normal distribution, and thus converts a proportion to a z-value under a normal assumption.

The weight $w_{\tau}$ specifies the weight that we assign to observations at iteration $\tau$. We assign higher weights to iterations closer to iteration $t$. A suggestion is to let the weights follow a normal distribution with mean $t$. The standard deviation of this normal distribution is a hyper parameter that we need to set. Higher values of the standard deviation of $w$ give more rigid growth curves. For the chess data used in this paper, a standard deviation of 2 works well.

As the goal is to estimate $\mu_t, \sigma_t$, which define the normal distribution at iteration $t$, we maximize expression \ref{eqn:likelihood_function} with regards to the mean and the standard deviation:
\begin{equation}
\label{eqn:likelihood_function_argmax}
\hat{\mu}_t, \hat{\sigma}_t = \underset{\mu_t, \sigma_t}{\text{argmax }} \prod_{\tau=1}^{n_{\tau}} \prod_{j=1}^{n_j}  w_{\tau} \cdot \cfrac{(e^{\hat{\theta}_{j,\tau}-d_{j,\tau}})^{y_{j,\tau}}}{1 + e^{\hat{\theta}_{j,\tau}-d_{j,\tau}}},\;\;\;\;\; \hat{\theta}_{j,\tau} = \hat{\mu}_{t} + \hat\sigma_{t} \cdot \Phi^{-1} \left( P_{j,\tau} \right).
\end{equation}

\end{enumerate}

\subsubsection{Estimate the ability of a new respondent}

Estimating the ability of a new respondent entails the following steps:

\begin{enumerate}
\item Calculate the Elo estimates for all iterations up until the most recent response.
\item For iteration $t$, calculate the proportion of respondents $P_{j,t}$ in the training data with lower ability estimates than the new respondent.
\item The estimated ability of respondent $j$ at iteration $t$ is calculated $\hat{\theta}_{j,t} = \hat{\mu}_{t} + \hat\sigma_{t} \cdot \Phi^{-1} \left( P_{j,t} \right)$.

\end{enumerate}

\subsubsection{Assumptions about distributional shapes}

In Item Response Theory we often make an initial assumption that the respondents in our population, as well as in our sample, have abilities that follow a standard normal distribution. We do not make that assumption here. However, we assume that all respondents who are at the same iteration $t$ have abilities that follow a normal distribution with unknown mean and unknown variance. That is, we treat the respondents who are answering their $t$:th item as one group. If there is reason to believe that the abilities for a time point follow a distribution other then the normal distribution, then this general method is still applicable.

\section{Data}

We want to mimic the situation of a group of learning students who regularly solve test items, e.g., provided by an ITS. Our aim is to estimate the growing ability of the students with good accuracy. For this purpose, we use data from young chess players, who typically improve their rating when playing regularly, especially when they belong to the group of higher rated players.

In a chess game, we know the rating as is was before the match for the player whose ability we want to track. We also know the rating of the opponent and the outcome. In an educational analogy, the rating of the player corresponds to the ability of a student. The rating of the opponent corresponds to the item difficulty. A game won corresponds to an item answered correctly, and a loss to an item answered incorrectly.

The International Chess Federation (FIDE) rates a large number of chess players based on their performance in FIDE rated tournaments.
The FIDE website lists all games that have been played within the FIDE rating system. The games are grouped by month. The chess rating data used for this paper were collected from the International Chess Federation FIDE in November 2023 \citep{fide_fide_2023}.

As we are interested in tracking the progress of players with changing abilities, we have collected game results from relatively young players born between 2000 and 2008. We have further limited our data collection to players who had a ranking of at least 2000 at the time of data collection to  focus even more on players which typically are improving over time, i.e., show an ability growth.

In a first step we limited the data to players who have at least 70 active months. In a second step we removed players who did not have at least 50 active months with either a win or a loss. This can be the case for players with several active months where the only outcomes were draws. We were left with 919 players. For these players  we used data from the 50 first active months, which gave a balanced dataset.

 The observations that we use are the player's first game each active month, which may be a win or loss. We do not use games that resulted in a draw, as a draw does not have an analogy in an educational setting. A player typically does not play every month. The chronological time between games is therefore irregular and can span from less than a month to several months. This is analogous to students who may solve exercises at different time intervals.

In most instances, players who are active during a particular month play several games during that month. The FIDE rating is calculated based on the outcomes of all these games. We viewed this rating as the truth when we evaluated different methods for estimating chess-playing ability. It should be noted that just like our estimates, the rating is itself an estimate, calculated with the Elo method. However, the rating is based on a considerably larger number of games than our own estimates. Thus, when we evaluate our method we examine how well it reproduces an estimate while using fewer data points.

In our dataset, 66 percent of the recorded games are one month or less apart. 98 percent are 6 months or less apart. The mean number of games that a player plays during an active month is 12, and the median is 10. The minimum is 1 and the maximum 109. The data was divided into a training set with 644 observations and a test set with 275 observations, since we decided to set aside 30 percent of the observations as test data. These 30 percent of the observations were randomly chosen.

Since we will use a model in the logistic form
\begin{equation}
\label{eqn:elo_res3}
P(Y=1) = \cfrac{e^{\theta - d}}{1 + e^{\theta - d}},
\end{equation}
we have transformed the ratings, i.e. the ability estimates, using the formula
\begin{equation}
\label{eqn:elo_res4}
\theta = \cfrac{R - 1500}{400} \cdot \ln(10).
\end{equation}

As the probability of a win depends on the difference in ability, subtracting 1500 is not necessary. However, we have done it to get initial abilities that are in the neighborhood of 0.

\section{Results}

In this section we compare our proposed method to other methods that can be appropriate for tracking changing abilities. We will consider the following methods.

\begin{itemize}
\item \textbf{GLMM fixed:} This is a generalized mixed model, but we only use the  fixed effects to estimate abilities. Thus all respondents get the same estimate at one iteration. These are the estimates we get from the R \textit{predict} function with an lme4 model.
\item \textbf{GLMM refit:} This is a GLMM model where we use both the fixed effects and the random effects for ability estimation. For each new estimate, the entire model was refit with observations for the new respondent included. For instance, to estimate the abilities of a new respondent at iteration 5, new data for iterations 1 to 5 are included, whereas data for subsequent iterations are missing. Observations with all iterations from the training data were included.
\item \textbf{GLMM ML:} Here we fit the GLMM model with the training data. We then estimate the ability of a new respondent by maximizing a custom likelihood function. The difference between this and \textit{GLMM refit} is that here the estimated model parameters are not influenced by the observations in the test data. In reality, the estimates from this method are almost identical that those of \textit{GLMM refit}, but it takes only a small fraction of the time to estimate new respondents. As \textit{GLMM ML} is clearly better than than \textit{GLMM fixed} and more practical than \textit{GLMM refit}, this will be the GLMM approach that we focus on mostly.
\item \textbf{Elo:} This is the standard Elo method, where we use an optimized constant step size.
\item \textbf{the Elo-informed growth model:} In this method, we rank respondents based on Elo estimates, and then we estimate individual abilities based on the assumption that the abilities at each iteration follow a normal distribution.
\end{itemize}

We were primarily interested in the accuracy and the speed of the different methods, but we also examined how much training data they require to work well. Some models described in this paper are not included in the comparison. This is because they do not meet one or more criteria. Learning factors analysis (LFA) and Performance factor analysis (PFA) are excluded because of their rigidity. The methods assume linear growth, as a function of the number of responses in the case of LFA and as a function of the number of correct and incorrect answers in the case of PFA. This assumption limits their usability to measuring progress that is reasonably linear, which may not always be the case. The Glicko method does not assume linearity, but it assumes that we improve the precision of the estimate after each measurement. This implies a lower variance of the prior for the next ability update, which results in smaller changes in the ability estimate. With a relatively flat ability over time, this logic is sound, but it is not appropriate when we assume ability growth that may be rapid. Knowledge tracing (KT) has the limitation that it does not account for different item difficulties, and is excluded from the comparison for that reason.

\subsection{Two test scenarios}

One purpose of this method comparison is to determine how well each method handles rapid ability change. Rapid changes are more difficult to track, especially if they do not follow a simple pattern, such as a straight line. In the scenario that we call scenario 1, we have access to one chess game from every active month. An active month is a month where the player has played, and either won or lost, at least one game. In total, we have data from 50 active months for each player, i.e. we have 50 iterations.

In our other scenario, which we call scenario 2, we only have access to one game from every other active month. This implies that we can expect the ability to have changed more from one iteration to the next. When we use one observation from every other active month, we have a total of 25 iterations per player.

\subsection{Identify appropriate values for the Elo step-size parameters}
\label{sec:choosing-k}

In our model comparison, the Elo algorithm is used for two of the methods, the standard Elo method and the Elo-informed growth model, which ranks respondents based on the Elo algorithm. The use of the the Elo algorithm requires that we decide on the step size $K$ in formula \ref{eqn:elo3}, which is the purpose of this preliminary results section.

The optimal step size is a trade-off between lag and volatility. When the ability is changing quickly, a small $K$ will result in an estimate that lags as the updates are too small to keep up with the changes in the true ability. Thus, even though the Elo method does not assume a specific shape of the growth curve, it does assume that the change is not too steep. A larger $K$ allows the estimate to find the approximate true ability more quickly, but at the price of larger volatility. Several papers have discussed the proper value of $K$ to achieve good Elo estimates \citep{antal_use_2013, papousek_adaptive_2014, niznan_student_2015, pelanek_applications_2016}. The hyper parameter value $K=0.4$ has been considered appropriate in some educational settings \citep{antal_use_2013, wauters_item_2012}.

When we use the standard Elo method, the goal is to achieve Elo estimates as close to the true abilities as possible. However, when we use the Elo algorithm in the Elo-infomed growth model only to rank respondents, we are instead interested in finding the value of $K$ that gives the most accurate ranking.

As a preliminary to the model evaluation, we explored whether the step size that gives the best Elo estimate is also the step size that gives the best result for the Elo-informed growth model. We tried all step sizes 0.1, 0.2, ..., 1.5. The RMSE calculated for each step size is the mean RMSE over all iterations. At this stage, we calculated the RMSE on the training data and not on the test data. As $K$ is a hyperparameter that is used in the subsequent comparison between methods, we do not want to set $K$ based on the test data.

\begin{figure}
\begin{center}
\includegraphics[scale=0.7]{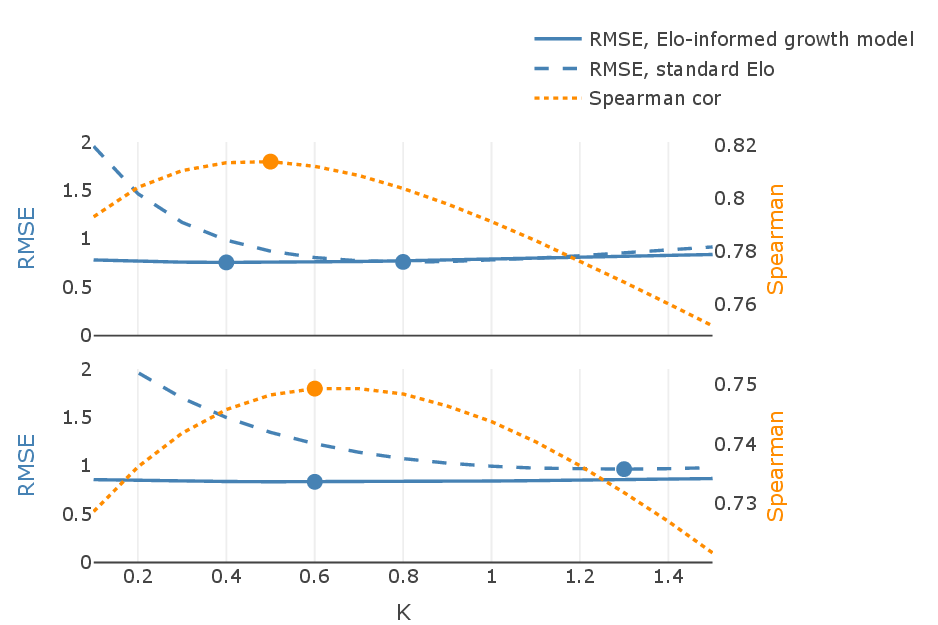
}
\caption{The plots show the RMSE for the Elo method, the RMSE for the Elo-informed growth model and the mean Spearman correlation as a function of the step size parameter $K$. The points mark the maxima and the minima of the curves. The plot shows that the optimal $K$ for the Elo-informed growth model is substantially smaller than the optimal $K$ for standard Elo estimation. The $K$ that maximizes the Spearman correlation is slightly larger or equal to the optimal $K$ for the Elo-informed growth model. Top: Test scenario 1. Bottom: Test scenario 2.}
\label{fig:k_effects}
\end{center}
\end{figure}

From figure \ref{fig:k_effects} we can see that the step size that gives the best Elo estimate is not optimal for the Elo-informed growth model. In scenario 1, the Elo estimate is most accurate with the step size that is approximately $K=0.8$. The step size that optimizes the accuracy for the Elo-informed growth model is $K=0.4$. Figure \ref{fig:k_effects} also shows that $K=0.5$ maximizes the Spearman correlation between the Elo estimates and true abilities. As the Spearman correlation measures the similarity between rankings \citep{spearman_proof_1904}, we would expect that the step size that maximizes the accuracy of the Elo-informed growth model is equal to or close to the step size that maximizes the Spearman correlation.

In scenario 2, we see a similar pattern. The optimal step size to achieve a good Elo estimate has now increased substantially to $K=1.3$. The optimal step size for the Elo-informed growth model has only increased slightly to $K=0.6$. The size that maximizes the Spearman correlation between the Elo ability estimates and the true abilities is also $K=0.6$.

The main take away from these results is that for the Elo-informed growth model we should not rely on values of $K$ that are chosen to give accurate Elo ability estimates. For data similar to the chess data used here, $K=0.5$ appears to be a step size that is quite robust regardless of the steepness of the ability growth. We can further note from figure \ref{fig:k_effects} that an inappropriate step size is of less consequence with the Elo-informed growth model compared to when we use the standard Elo estimate.

Based on these first results, in the following sections we use the step size $K=0.8$ for the Elo method and $K=0.4$ for the Elo-informed growth model for test scenario 1. For test scenario 2, we use $K=1.3$ for the Elo method and $K=0.6$ for the Elo-informed growth model.

\subsection{Compare model accuracy}

We compared the Elo-informed growth model with three other methods, in both scenario 1 and scenario 2. In each scenario, we fitted a model with the training observations and evaluated the model on the test observations.

\begin{figure}
\begin{center}
\includegraphics[scale=0.7]{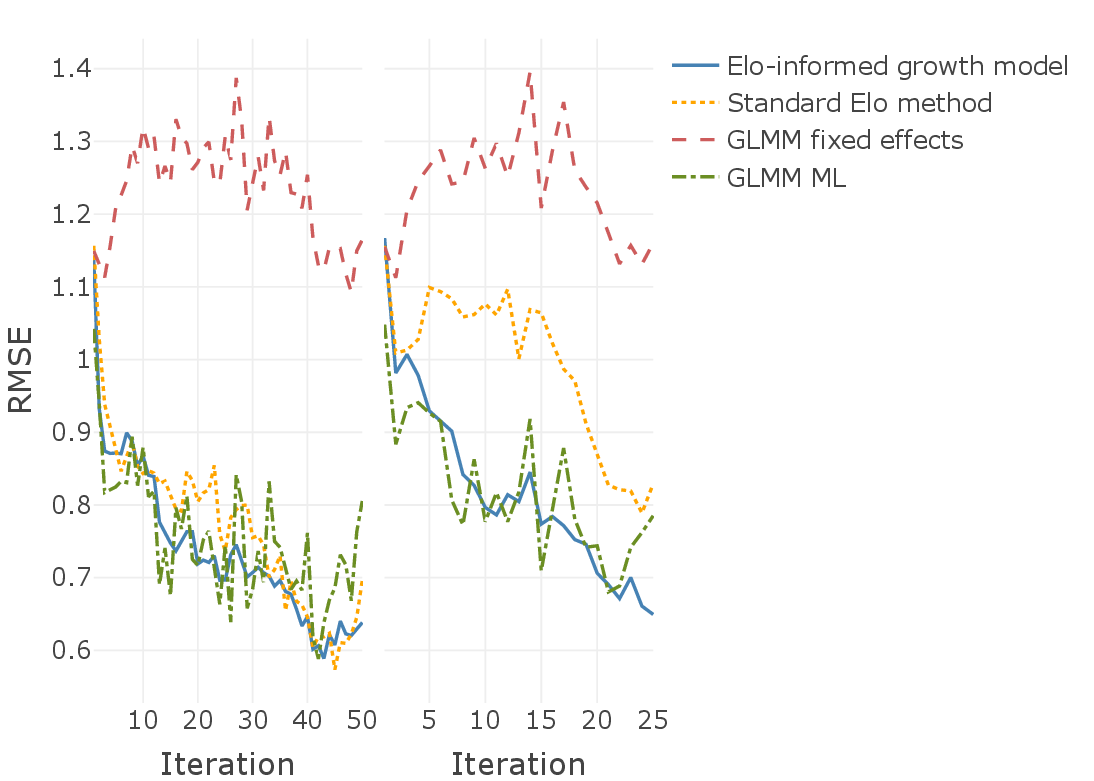}
\caption{Left: In test scenario 1, the Elo-informed Growth model performs similar to the Elo method. Right: In test scenario 2, the Elo-informed growth model perform better than the Elo method.}
\label{fig:RMSE_chess}
\end{center}
\end{figure}

Figure \ref{fig:RMSE_chess} shows that in scenario 1, the Elo-informed growth model performs similar to the standard Elo method and to the GLMM ML method. In scenario 2, the Elo-informed growth model and GLMM ML still perform equally well, while the standard Elo method performs somewhat worse. The GLMM fixed effects method performs badly in both scenarios.

\begin{figure}
\begin{center}
\includegraphics[scale=0.7]{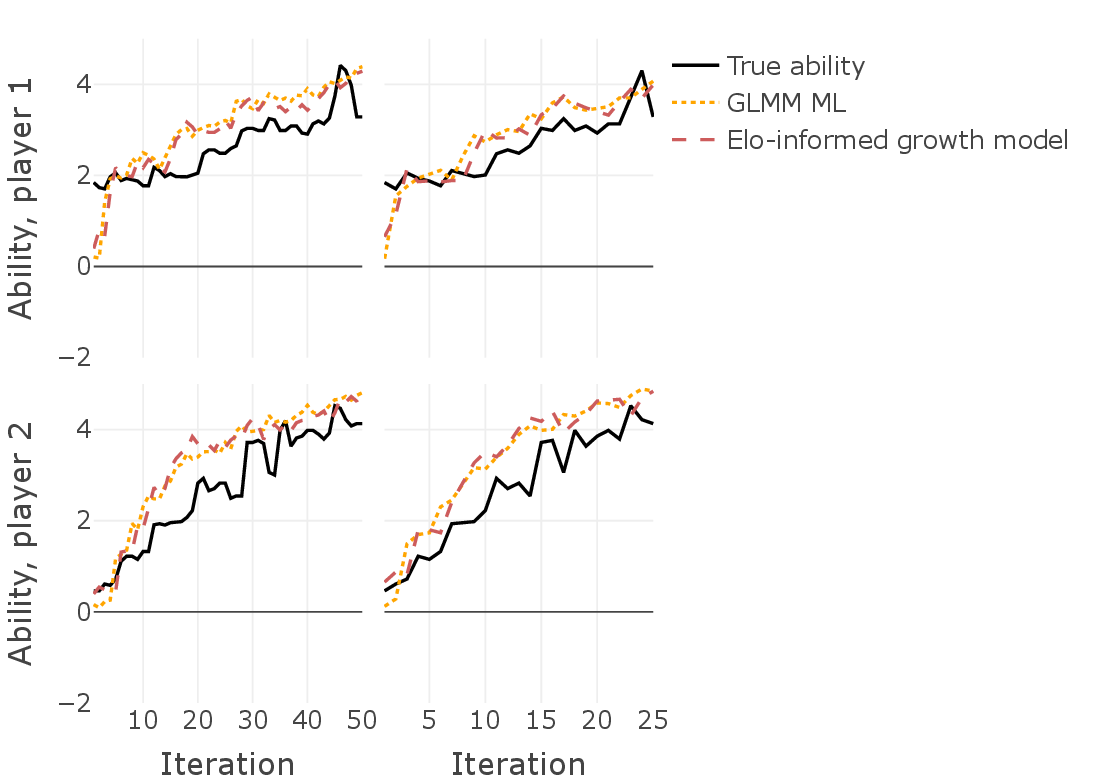
}
\caption{Left: Two examples of respondent ability growth in the test dataset in test scenario 1. We see the true ability (i.e the official rating), the GLMM ML estimates and the estimates with the Elo-informed growth model. Right: We have the same examples as on the left, but for test scenario 2.}
\label{fig:ability_examples}
\end{center}
\end{figure}

\begin{figure}
\begin{center}
\includegraphics[scale=0.7]{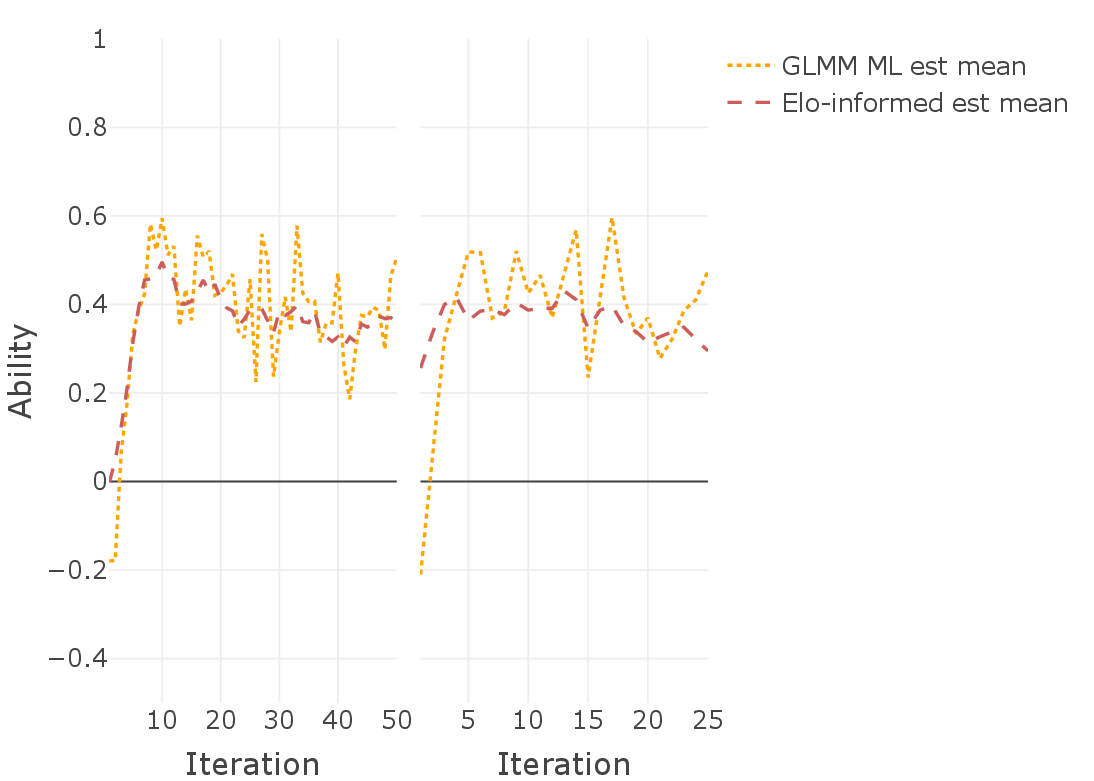
}
\caption{The figures show the means of the estimated abilities with the Elo-informed model minus the mean of the true abilities, and the mean of the GLMM ML model minus the mean of the true abilities.Left: Scenario 1 with an observation per active month. Right: Scenario 2 with an observation per every other active month.}
\label{fig:ability_means}
\end{center}
\end{figure}

Figure \ref{fig:ability_examples} shows the ability growth for two randomly chosen chess players. Each row in the plot represents one player, and each column represents one scenario. The left-hand column represents scenario 1, and the right-hand column represents scenario 2. For both players we see that the Elo-informed growth model and the GLMM ML produce very similar estimates. We note that the estimates are consistently higher than those implied by the official rating. This is confirmed when we look at the ability means in figure \ref{fig:ability_means}, where we can see that the estimates typically overestimate the true abilities with approximately 0.4. We should recall that the abilities that we consider to be true are also estimates, albeit based on a larger number of played games.

\subsection{Model properties}

Table \ref{tab:compare_summary} compares the models that we have evaluated. For the comparisons, with the exception of the comparison of different training data set sizes, we used a training set with 643 chess players over 50 iterations. Each iteration represents a month in which the player participated in at least one FIDE rated game. To measure the time it took to fit models and to estimate individual abilities, we used a PC with an Intel Core i7-10700K CPU at 3.80GHz and 16GB RAM. The processing was done in R version 4.4.0, and for mixed models we used the package \textit{lme4}, version 1.1-35.3.

The first row in table \ref{tab:compare_summary} shows that the GLMM models required more than 20 minutes to fit the model parameters. In contrast, the Elo algorithm requires virtually no time to train the model. As implemented here, training the Elo method means that we find good starting value for the algorithm based on the training data. If we already have an idea about a good starting value, the Elo method needs no model training. the Elo-informed growth model requires a little more than 10 seconds to fit the model parameters, i.e. to set a mean and a variance that define the distribution of abilities at each iteration. A reason for the speed difference between the proposed method and the GLMM model is that the GLMM model fits the means for all iterations along with the ability variance jointly. The Elo-informed growth model fits each iteration individually, which makes it a simpler optimization problem. However, note that whereas the distributions at the iterations are fitted individually, the distribution at each iteration takes the data from all iterations into account.

The second row in table \ref{tab:compare_summary} gives the speed of estimating the ability of a new player once the model has been fit. The GLMM refit method, which estimates a new respondent by refitting the model with the new observations included, takes as much time as fitting the original model. The GLMM fixed effects method estimates a new respondent using only the fixed effects, which is very quick as it requires no iterative procedure. With the GLMM ML method we find the random effect for the new respondent through a fast optimization procedure with only one parameter to estimate. The Elo method is updated with a simple formula with takes virtually no time to calculate. The Elo-informed growth model is slower than the standard Elo method, but fast enough for any practical application.

The third row specifies whether a method requires an iterative procedure to estimate or update the ability of a new respondent. An iterative procedure refers to a procedure where the estimate is found by numerical optimization of a function such as a maximum likelihood function. Typically, the need for numerical optimization implies a slower process, but as in the case of the GLMM ML method a simple optimization of a scalar may be fast.

Accuracy on rows four and five refers to how close the estimated abilities are to the players' true ratings in each of the two test scenarios. Slower change corresponds to scenario 1, and rapid change corresponds to scenario 2. The GLMM methods that account for random effects in the estimates perform well, just as the Elo-informed growth model. When we use GLMM and estimate abilities based only on the fixed effects, we get bad accuracy. The standard Elo method does well when the ability changes at a slower pace, but has problems following more rapid change. 

\begin{figure}
\begin{center}
\includegraphics[scale=0.7]{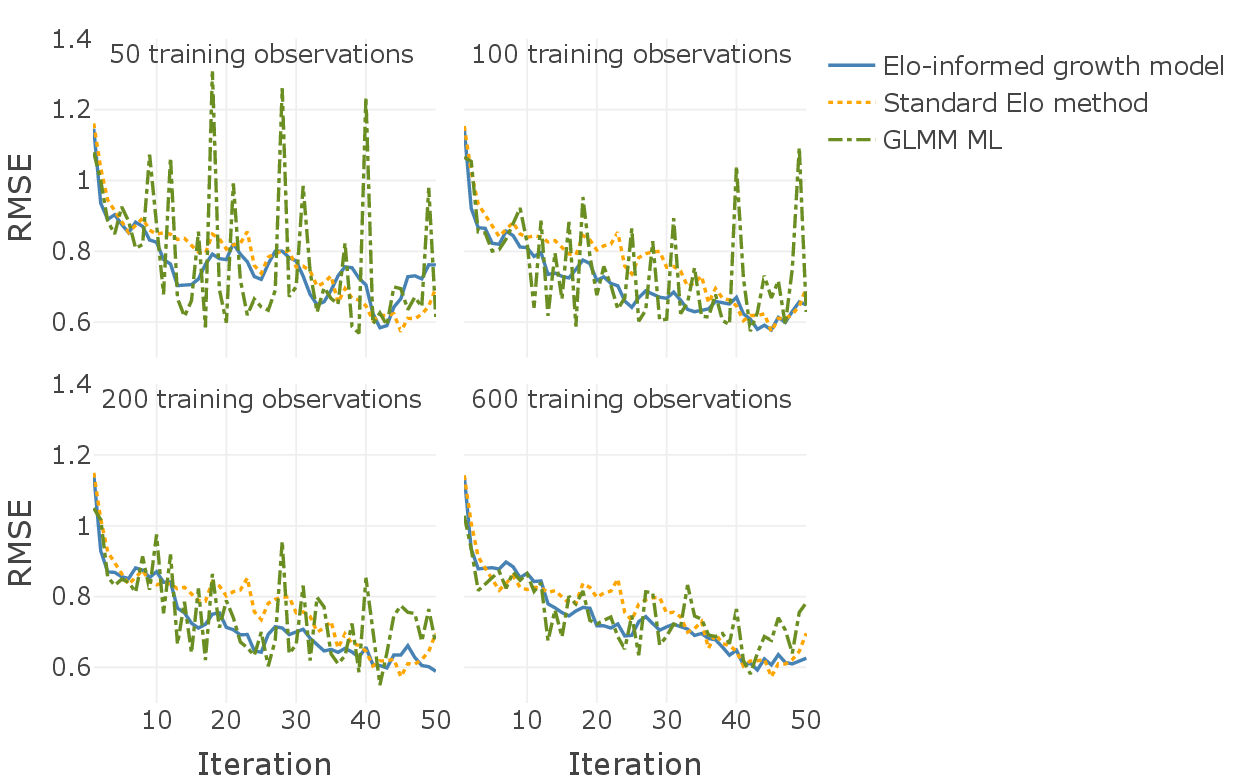}
\caption{RMSE with different number of observations in the training dataset.}
\label{fig:training_size}
\end{center}
\end{figure}

The last row shows how many observations that are needed in the training data to get reasonably good and stable estimates. As we see in figure \ref{fig:training_size}, the GLMM models need around 600 observations in the training data to produce reasonably good and stable estimates. The Elo-informed growth model gives good estimates with only about 50 training observations. The standard Elo method does not need any training data as long as we can make a reasonable guess of the initial ability as a starting point for the algorithm.

\begin{table}[ht]
\centering
\begin{tabular}{llllll}
  \hline
 & Elo-informed & Elo & GLMM fixed effects & GLMM refit & GLMM ML \\ 
  \hline
Model fitting speed & 11s & 0.01s & 23m 16s & 23m 16s & 23m 16s \\ 
  Theta fitting speed & 0.02s & $<$ 0.01s & 0.01s & 23m 16s* & 0.01s \\ 
  Iterative theta fitting & No & No & No & Yes & Yes \\ 
  Accuracy, slower change & Good & Good & Bad & Good & Good \\ 
  Accuracy, faster change & Good & Average & Bad & Good & Good \\ 
  Min training obs & 50 & 0** & 600 & 600 & 600 \\ 
  \end{tabular}
\caption{The table compares several properties of methods for estimating abilities. The vaues are calculated based on 643 observations and 50 iterations in the training dataset. *Estimating the ability of a new respondent entails a refit of the model with one additional observation, which takes as long as fitting the original model. **The Elo method does not need training data. However, a small amount of training data can help with the specification of a good starting value for the algorithm.} 
\label{tab:compare_summary}
\end{table}

\section{Conclusion and discussion}

Several methods have been developed to track the changing ability of a respondent. These methods often assume a certain shape of the growth curve, e.g. a growth that is linear. The GLMM model with time as a factor variable does not assume a shape of the ability growth curve, and neither does the Elo method. However, GLMM models of this type are slow to fit, and the Elo method will tend to underestimate the changes in ability when we use a small step size. Increasing the step size may not improve the estimates, since a larger step size leads to more volatile ability estimates.

In this paper, we propose the Elo-informed growth model for tracking changing abilities. This method does not assume a distributional shape. The method uses the Elo algorithm to rank respondents, but unlike standard Elo estimates, the estimates from our proposed method are robust to rapid ability changes.

When we evaluate our method on real chess data, we can see that see that our method has a performance similar to that of the Elo algorithm when we use more frequent measurements. However, our method performs better than the Elo algorithm when we use less frequent measurements. The reason for this is that less frequent measurements imply more rapid change in ability between the measurements. The Elo-informed growth model has a similar performance as that of a GLMM model where we use both fixed effects and random effects for the ability estimation. However, the proposed model is considerably faster to fit. There are several differences in the assumptions when we compare the proposed method and the specified GLMM methods.

\begin{enumerate}
\item The between-player variance of abilities is constant in the GLMM model that we defined, although in theory it could vary at the cost of further complexity. In the Elo-informed growth model, it is allowed to vary between iterations.
\item If player 1 is estimated to be better than player 2 at some iteration, that implies that player 1 is estimated to be better at all iterations in the GLMM models. In contrast, the Elo-informed growth model allows players to switch positions in the ranking from one iteration to the next.
\item The GLMM models fit the model parameters jointly. The Elo-informed growth model fits the model parameters for each iteration separately. However, when it fits the parameters for one iteration, it takes into account also the data from other iterations.
\end{enumerate}

The Elo-informed growth model is useful under the assumption that we have access to calibrated items, i.e. items of known difficulty. We also assume unidimensionality, so that an item measures only one ability. The method is especially useful when we either expect rapid ability changes or when we have no or little prior knowledge about the shapes of the ability growth curves.

\bibliographystyle{apa-good.bst} 
\bibliography{draft2_references.bib}

\section{Appendix}

Figure \ref{fig:RMSE_chess} shows RMSE for the estimated ability, calculated with different estimation methods. Whereas figure \ref{fig:RMSE_chess} only includes the methods considered most useful, table \ref{tab:RMSE_comparison_50} and table \ref{tab:RMSE_comparison_25} include all methods that are compared in table \ref{tab:compare_summary}.

\begin{table}[ht]
\centering
\begin{tabular}{cccccc}
  \hline
 & \multicolumn{5}{c}{Rooted Mean Squared Error (RMSE)}\\
 \cmidrule(lr){2-6}Iteration & Elo-informed & Elo & GLMM fixed effects & GLMM refit & GLMM ML \\ 
  \hline
  1 & 1.14 & 1.16 & 1.15 & 1.05 & 1.04 \\ 
    2 & 0.93 & 1.03 & 1.13 & 0.95 & 0.94 \\ 
    3 & 0.87 & 0.94 & 1.11 & 0.82 & 0.82 \\ 
    4 & 0.87 & 0.91 & 1.16 & 0.82 & 0.82 \\ 
    5 & 0.87 & 0.88 & 1.21 & 0.81 & 0.82 \\ 
    6 & 0.87 & 0.85 & 1.23 & 0.81 & 0.83 \\ 
    7 & 0.90 & 0.87 & 1.25 & 0.78 & 0.83 \\ 
    8 & 0.89 & 0.89 & 1.29 & 0.87 & 0.89 \\ 
    9 & 0.85 & 0.85 & 1.27 & 0.82 & 0.83 \\ 
   10 & 0.87 & 0.84 & 1.32 & 0.86 & 0.88 \\ 
   11 & 0.84 & 0.85 & 1.29 & 0.78 & 0.81 \\ 
   12 & 0.84 & 0.84 & 1.31 & 0.76 & 0.82 \\ 
   13 & 0.78 & 0.83 & 1.24 & 0.68 & 0.69 \\ 
   14 & 0.76 & 0.83 & 1.27 & 0.77 & 0.74 \\ 
   15 & 0.75 & 0.81 & 1.24 & 0.68 & 0.68 \\ 
   16 & 0.74 & 0.79 & 1.33 & 0.82 & 0.80 \\ 
   17 & 0.75 & 0.79 & 1.31 & 0.79 & 0.77 \\ 
   18 & 0.76 & 0.85 & 1.30 & 0.88 & 0.81 \\ 
   19 & 0.76 & 0.83 & 1.26 & 0.70 & 0.73 \\ 
   20 & 0.72 & 0.80 & 1.27 & 0.68 & 0.72 \\ 
   21 & 0.72 & 0.82 & 1.29 & 0.74 & 0.75 \\ 
   22 & 0.72 & 0.82 & 1.30 & 0.76 & 0.76 \\ 
   23 & 0.73 & 0.85 & 1.24 & 0.74 & 0.71 \\ 
   24 & 0.70 & 0.76 & 1.24 & 0.68 & 0.66 \\ 
   25 & 0.70 & 0.74 & 1.31 & 0.74 & 0.74 \\ 
   26 & 0.73 & 0.78 & 1.27 & 0.66 & 0.64 \\ 
   27 & 0.75 & 0.79 & 1.39 & 0.80 & 0.84 \\ 
   28 & 0.72 & 0.80 & 1.33 & 0.85 & 0.80 \\ 
   29 & 0.70 & 0.80 & 1.21 & 0.65 & 0.66 \\ 
   30 & 0.71 & 0.76 & 1.24 & 0.68 & 0.68 \\ 
   31 & 0.71 & 0.76 & 1.28 & 0.70 & 0.74 \\ 
   32 & 0.71 & 0.74 & 1.23 & 0.65 & 0.69 \\ 
   33 & 0.70 & 0.70 & 1.33 & 0.75 & 0.83 \\ 
   34 & 0.69 & 0.71 & 1.27 & 0.74 & 0.75 \\ 
   35 & 0.70 & 0.73 & 1.25 & 0.82 & 0.74 \\ 
   36 & 0.68 & 0.65 & 1.29 & 0.72 & 0.71 \\ 
   37 & 0.68 & 0.69 & 1.23 & 0.74 & 0.68 \\ 
   38 & 0.66 & 0.67 & 1.23 & 0.73 & 0.70 \\ 
   39 & 0.63 & 0.66 & 1.21 & 0.69 & 0.68 \\ 
   40 & 0.65 & 0.65 & 1.25 & 0.76 & 0.76 \\ 
   41 & 0.60 & 0.60 & 1.17 & 0.64 & 0.62 \\ 
   42 & 0.61 & 0.62 & 1.13 & 0.60 & 0.59 \\ 
   43 & 0.59 & 0.62 & 1.12 & 0.67 & 0.64 \\ 
   44 & 0.62 & 0.62 & 1.15 & 0.63 & 0.67 \\ 
   45 & 0.61 & 0.57 & 1.15 & 0.67 & 0.69 \\ 
   46 & 0.64 & 0.61 & 1.15 & 0.76 & 0.73 \\ 
   47 & 0.62 & 0.61 & 1.12 & 0.71 & 0.72 \\ 
   48 & 0.62 & 0.62 & 1.09 & 0.66 & 0.67 \\ 
   49 & 0.63 & 0.65 & 1.15 & 0.71 & 0.76 \\ 
   50 & 0.64 & 0.70 & 1.16 & 0.80 & 0.81 \\ 
  \end{tabular}
\caption{The Rooted Mean Squared Error (RMSE) for each method evaluated after 1, 2, ..., 50 iterations. The models are based on one observation for each active month, i.e. one observation for each month where a player has played at least one rated game.} 
\label{tab:RMSE_comparison_50}
\end{table}

\begin{table}[ht]
\centering
\begin{tabular}{cccccc}
  \hline
 & \multicolumn{5}{c}{Rooted Mean Squared Error (RMSE)}\\
 \cmidrule(lr){2-6}Iteration & Elo-informed & Elo & GLMM fixed effects & GLMM refit & GLMM ML \\ 
  \hline
  1 & 1.17 & 1.16 & 1.15 & 1.07 & 1.05 \\ 
    2 & 0.98 & 1.01 & 1.11 & 0.88 & 0.88 \\ 
    3 & 1.01 & 1.01 & 1.21 & 0.92 & 0.93 \\ 
    4 & 0.98 & 1.03 & 1.25 & 0.89 & 0.94 \\ 
    5 & 0.93 & 1.10 & 1.27 & 0.92 & 0.93 \\ 
    6 & 0.92 & 1.09 & 1.29 & 0.89 & 0.92 \\ 
    7 & 0.90 & 1.08 & 1.24 & 0.80 & 0.81 \\ 
    8 & 0.84 & 1.06 & 1.24 & 0.77 & 0.77 \\ 
    9 & 0.83 & 1.06 & 1.30 & 0.89 & 0.86 \\ 
   10 & 0.80 & 1.08 & 1.26 & 0.75 & 0.78 \\ 
   11 & 0.79 & 1.06 & 1.30 & 0.81 & 0.82 \\ 
   12 & 0.81 & 1.10 & 1.25 & 0.80 & 0.78 \\ 
   13 & 0.80 & 1.00 & 1.31 & 0.82 & 0.82 \\ 
   14 & 0.85 & 1.07 & 1.40 & 0.88 & 0.92 \\ 
   15 & 0.77 & 1.06 & 1.21 & 0.71 & 0.71 \\ 
   16 & 0.78 & 1.02 & 1.29 & 0.76 & 0.79 \\ 
   17 & 0.77 & 0.99 & 1.35 & 0.80 & 0.88 \\ 
   18 & 0.75 & 0.97 & 1.26 & 0.87 & 0.78 \\ 
   19 & 0.75 & 0.91 & 1.24 & 0.80 & 0.74 \\ 
   20 & 0.71 & 0.87 & 1.22 & 0.75 & 0.74 \\ 
   21 & 0.69 & 0.83 & 1.17 & 0.70 & 0.68 \\ 
   22 & 0.67 & 0.82 & 1.13 & 0.72 & 0.69 \\ 
   23 & 0.70 & 0.82 & 1.16 & 0.73 & 0.74 \\ 
   24 & 0.66 & 0.79 & 1.13 & 0.75 & 0.76 \\ 
   25 & 0.65 & 0.83 & 1.16 & 0.73 & 0.79 \\ 
  \end{tabular}
\caption{The Rooted Mean Squared Error (RMSE) for each method evaluated after 1, 2, ..., 25 iterations. The models are based on one observation for every other active month, i.e. one observation for every other month where a player has played at least one rated game.} 
\label{tab:RMSE_comparison_25}
\end{table}

\end{document}